\documentclass[11pt]{article}
\usepackage[utf8]{inputenc}
\usepackage{jheppub}
\usepackage{amsmath}
\usepackage{physics}
\usepackage{tikz}
\usepackage{cleveref}
\usepackage{comment}
\usepackage{natbib}

\usepackage{subfigure} 

\usepackage{graphics,epsfig}
\usepackage{graphicx}

\usepackage{amsmath}
\usepackage{amssymb}
\usepackage{bbm}
\usepackage{subfigure}
\usepackage{epsfig}
\usepackage{yfonts}
\newcommand{\bea}{\begin{eqnarray}}
\newcommand{\eea}{\end{eqnarray}}
\newcommand{\bean}{\begin{eqnarray*}}
\newcommand{\eean}{\end{eqnarray*}}

\renewcommand{\H}{\mathcal{H}}

\newcommand{\M}{\mathcal{M}}
\newcommand{\e}{\epsilon}

\newcommand{\be}{\begin{eqnarray}}
\newcommand{\ee}{\end{eqnarray}}

\def\braket#1{\left\langle #1 \right\rangle}

\def\braket#1{\left\langle #1 \right\rangle}

\def\Tr{\mathop{\rm Tr}}

\title{Maxwell's Demon for Emergent Page Curve and Split Property}
\author[a,b]{Yang An}
\affiliation[a]{Institute for Theoretical Physics \& Cosmology, Zhejiang University of Technology, Hangzhou, 310023, China}
\affiliation[b]{United Center for Gravitational Wave Physics (UCGWP),
Zhejiang University of Technology, Hangzhou, 310023, China}
\emailAdd{anyangpeacefulocean@zju.edu.cn}
\abstract{

Seeking for the proper situation of Emergent Gravity, we recent reveal that the entropic mechanism only happens when extremal surfaces are varied, which is similar to the non-gravitational-bath-coupled setup of the island development.  
In this paper, we consider perturbing thin shell state outside horizon during equilibrating, to find the evolution of these Euclidean states requires an external force to be $ F_{ex}\propto T_{H}\delta A(\mu_a)$, proportional to area variation of the apparent horizon which could transform into the actual event horizon as extremal surface. 
We analogize the falling tendency to the 2nd law tendency, and then explore the potential tendency violation of Hawking radiation.

This analogy recalls the role of Maxwell's Demon to interpret the mechanism of the bath for the Page curve to emerge in a close universe. It could reconcile the Split Problem concerning quantum gravity raised by Raju.

}
\begin{document}
\maketitle
\section{Introduction}
The gravitational tendency could be a reflection of the 2nd law direction. In term of thermodynamics, the free falling tendency is just along the coarse-graining direction. How about the Hawking radiation into external non-gravitational systems?

Recent breakthrough \cite{Penington:2019npb,Almheiri:2019psf} towards understanding Black Hole Information Paradox (BHIP) is carrying on a special substance when a holographic gravity theory coupled to a non-gravitating bath. Based
 on the research of Quantum Extremal Surface (QES) \cite{Engelhardt:2014gca}, such bath setup is mean to make the QES as well as entanglement wedge bounded by QES no longer stay the same, to get the Page curve of the time-dependent holographic entanglement entropy pivoted. Computation on the  double holographic models was proposed in \cite{Almheiri:2019hni}, from simplicity of 2D dilation model based on JT gravity, which eventually leads to a significant triumph, the island rule \cite{Almheiri:2019hni}. Then a huge amount of development is carrying on this setup. They prompt our understanding of unitarity of quantum gravity theory, and confirm no information loss during black hole evaporation even in semi-classical level. 

Such special bath coupling is through modifying the AdS boundary condition from reflecting to absorbing \cite{Almheiri:2019psf}, intuitively opens the AdS box that keeps the black hole inside not evaporated out as the asymptotic flat black hole. 
 Therefore one get QES phase transition as well as entanglement wedge varying for evaporating black holes \cite{Penington:2019npb}. 

Through this setup, one could split the whole Hilbert space bipartitely into $H_{R}\otimes H_{bh}$, ``radiation" in the bath subsystem and ``black hole" in the gravitational subsystem, which is a realization of Page's argument \cite{Page:1993wv} from unitarity and general expectations of quantum information theory.  The disconnected island region \cite{Maldacena:2013xja} appears in the entanglement wedge of ``radiation" after Page time to get a phase transition of QES.
 Latter, the transition could be understand from the saddle point changing of Euclidean Path integral, that each state on time slice is a different Euclidean state contributed from Replica wormholes \cite{Penington:2019kki,Almheiri:2019qdq}.

 However, the Split Problem was raised in \cite{Raju:2021lwh} towards severely questioning adopting this fundamental intuitively split to a holographic theory: the Hilbert space of a single quantum gravity theory may not have the essential split property like ordinary quantum field theory. 
 This simple non-gravitating-bath-coupling breaks the gravitational Gauss's Law \cite{Geng:2021hlu}, such that the gravity theory suffers from massive graviton problem \cite{Geng:2020qvw} in these double holographic models.
 While, the holographic principle of gravity \cite{Susskind:1994vu,Bousso:2002ju} tells that the boundary already encodes all the information inside, which may be visualized as holographic screen with lower dimensional quantum field degree of freedom. Switching to a gravitating bath will lead to a trivial Page curve \cite{Geng:2020fxl}, the constant line rather than being pivoted. This trivial Page curve could also be the general property even for asymptotic flat holography with quantum dual on null infinity \cite{Laddha:2020kvp}.   
\subsection{Entropic Tendency from the Emergent Gravity development} 
The resolving of the puzzle also concerns another problem hanging on Emergent Gravity theories \cite{Verlinde:2010hp,Verlinde:2016toy,Jacobson:1995ab,Jacobson:2015hqa,Faulkner:2013ica}: should there be an entropic mechanism for gravity/gravitational force to emerge?
The similarity to the Page Curve is that they both share the problem on what causes the entropy variation.

Our inspiration comes from the question on how to distinguish the direction of falling tendency, as emergent phenomenon from  the thermodynamic 2nd Law. We take completely different consideration than the original entropic gravity theories to treat this question.
After that, we turn to generalize this consideration to gravitational tendency of Hawking radiation.

In \cite{An:2018hyt,An:2020ncr}, we developed a special method based on comparing the Entanglement First Law for the excited state and vacuum, to consider the entropy associated with gravitational attraction. We find the entropic variation happened when $F_g+F_{ex}=0$ to make a specific quasi-static process like a reversible heat engine. Locally it is an isoenergic process enduring a local Hawking temperature field linked by red-shift factor, that leads to the derivation of gravitational inertial force. The amount of entropy variation is exactly the variation of the Casini-Bekenstein bound \cite{Casini:2008cr} $\Delta S\leq\Delta \braket K$ and we may finally extend the re-derivation of Einstein Equation from thermodynamics \cite{Jacobson:1995ab,Verlinde:2010hp,Jacobson:2015hqa,Faulkner:2013ica} to generic situations more valid. 

Through such entropic mechanism, we are able to tell the falling tendency as thermodynamic tendency from variation of the entropy bound:
\bea
\bold F_g= -T_H\nabla_\mu \left(\frac{\delta Area(\Sigma_{r_s})}{4G}\right)
\eea
 where the entropy is exactly related to the variation of horizon $\Sigma_{r_s}$ area as the extremal surface $\gamma$ after black hole absorbs the test particle. For a consistent check, we notice from recent paper \cite{Engelhardt:2017aux,Engelhardt:2018kcs, Chandra:2022fwi} that,  its holographic meaning corresponds to  the apparent horizon $\gamma_a$.


In this paper, we will reconsider the former special method we called ``Perturbing on the Perturbation" technique for evolving excited state in quasi-equilibrium, and reinterpret this development of Emergent Gravity theory as the evolution of Euclidean states with the bath coupling.  And only in this special case to evolve euclidean state in quasi-static, the entropy associated with gravitational attraction coinsid to holographic entanglement entropy bound and coarse-grained entropy with a good holographic meaning
\bea
S(M,m)=\lim_{upper}\Delta S=S_\text{coarse}=\frac{\delta A(\mu_a)}{4G}\,.
\label{Sentcoar}
\eea

What's more, it also provides a way to tell the gravitational tendency from entropic variation for heat flux into the bath, just like the original Verlinde's Entropic Force Conjecture \cite{Verlinde:2010hp, Verlinde:2016toy} was meant to unify thermodynamic tendency and falling tendency.  We will explore to apply it to analyze the evaporation process of black holes.
 With the help of external auxiliary system, the coarse-graining direction, which is identity as the opposite to $\bold F_{ex}$, could indicate whether the evolving direction of black hole evaporation into the bath, before and after the Page time, is violating original gravitational tendency or not.

\subsection{Involve the Maxwell's Demon}

Here we remind the historical Maxwell's demon to cause a paradox towards the thermodynamic 2nd law. In this well-known thought experiment, Maxwell imagined a microscopic demon controlling a door to separate two kind of gases into two rooms. This demon can secretly distinguish the microscopic behavior of gas molecules and select some molecules to pass while keeps others not. In this way,  the system evolves against the disorder direction of entropy increasing, which seems to violate the thermodynamic 2nd law. 

The true explanation is that the demon's hidden information was secretly excluded for the whole system. While, after including back the memory it stores, the whole system still get recovered to follow the 2nd law. Modern development includes quantum Maxwell's demon \cite{Lloyd:1996wz}, quantum Maxwell's demon with non-Markvian effect \cite{Poulsen:2021erv}, discussion about the time arrow \cite{Yanik:2022snv}, various applications and experimental observation \cite{PhysRevLett.113.030601,Cottet_2017}. 

When we try to unify gravitational tendency with thermodynamic tendency, we could tell when or when not the system is evolving along the direction of tendency by taking entropic variation as the orientation. Even just in such specific situations of emergent gravity theory to explain falling tendency,  we have gotten the glimpse of such unification.

\begin{figure}[hbt]
		\centering
  \includegraphics[width=5in]{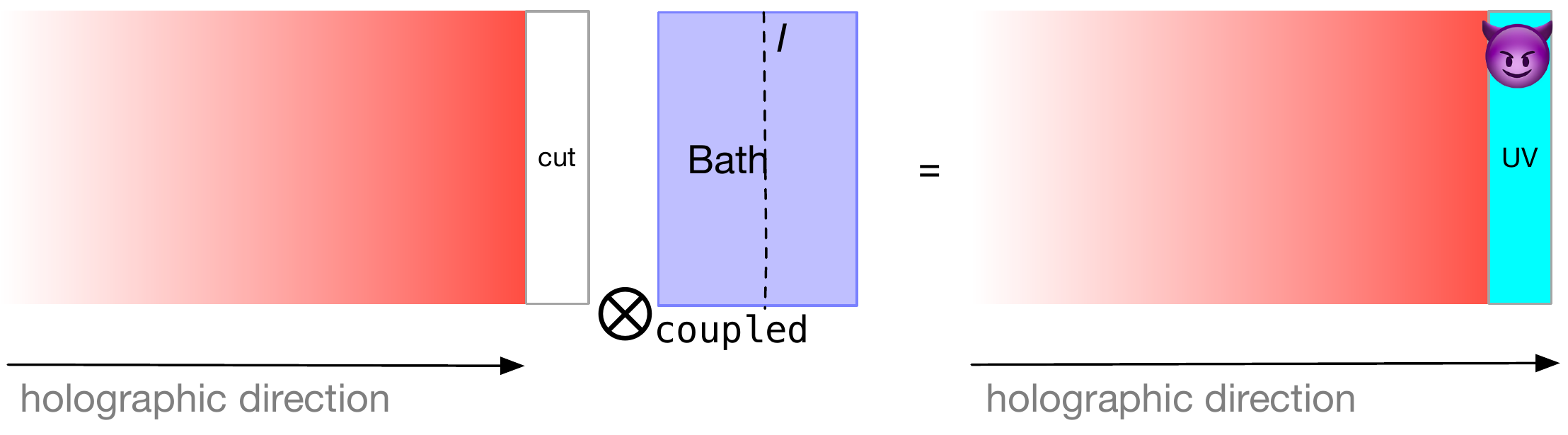}
  \caption{From an open system coupled to external bath, we propose it could be imitated by a close system with a felicity like Maxwell's Demon in the UV part.}
  \label{demoneq}
\end{figure}

The demon's role could be introduced for an interesting interpretation for relations between different entropy curves.  In the closed Quantum Gravity system like AdS/CFT, one could regard the role of the demon as the bath setup, to violate the ordinary AdS tendency of keeping the black hole in the box, such that the black hole evaporating into to other Euclidean states  otherwise  it shouldn't have done, to get the island saddle for a pivoted Page curve. 
The role of Maxwell's Demon in the closed QG system, could be a way to understand the real reason for the split property by the non-gravitating bath-coupling.  Two pictures could be equivalent to each other as in Figure \ref{demoneq}.

However,  when recover back this cutout of UV region, as if restore the hidden information by the demon, we are able to recover the whole quantum theory of gravity.  The unnoticeable demon could be as weak as the weak gravity towards radiation, but they shouldn't be ignorant for unitarity, therefore the holography and the gravitational Gauss's law get recovered for a trivial Page curve. Besides, if we regard the evaporation evolves not with the real time $t$ synchronically but with other parameter $\lambda$ in equilibrium, the absorbing boundary condition is just some simplification of device acts like the Maxwell's demon, so we will not get the massive graviton at all.


\section{Preliminary}
Suppose the decomposition of Hilbert space of a quantum system
(see a review for von Neumann Algebra with irreducible representation such as \cite{Witten:2018lha}), 
for any global state with density matrix $\rho=|\Psi\rangle\langle \Psi |$, the state confined in the subsystem $A$ (whose complement is $\bar{A}$) can be described by the reduced density matrix $\rho_A=\Tr_{\bar{A}} \rho$. We can always write the reduced density matrix as
\bea
\rho_A=\frac{e^{-K}}{\Tr{e^{-K}}}\,,
\eea
because it is positive defined and hermitian. $K$ is known as the modular Hamiltonian \cite{Haag:1992hx} of $\rho_A$. The entanglement entropy is defined as the von Neumann entropy
\bea
S_{vn}=S(\rho_A)=-\Tr\rho_A\log{\rho_A}\,.
\label{entS}
\eea

\subsection{Unitary Evaporation and QES}

\begin{figure}[hbt]
		\centering
  \includegraphics[width=5in]{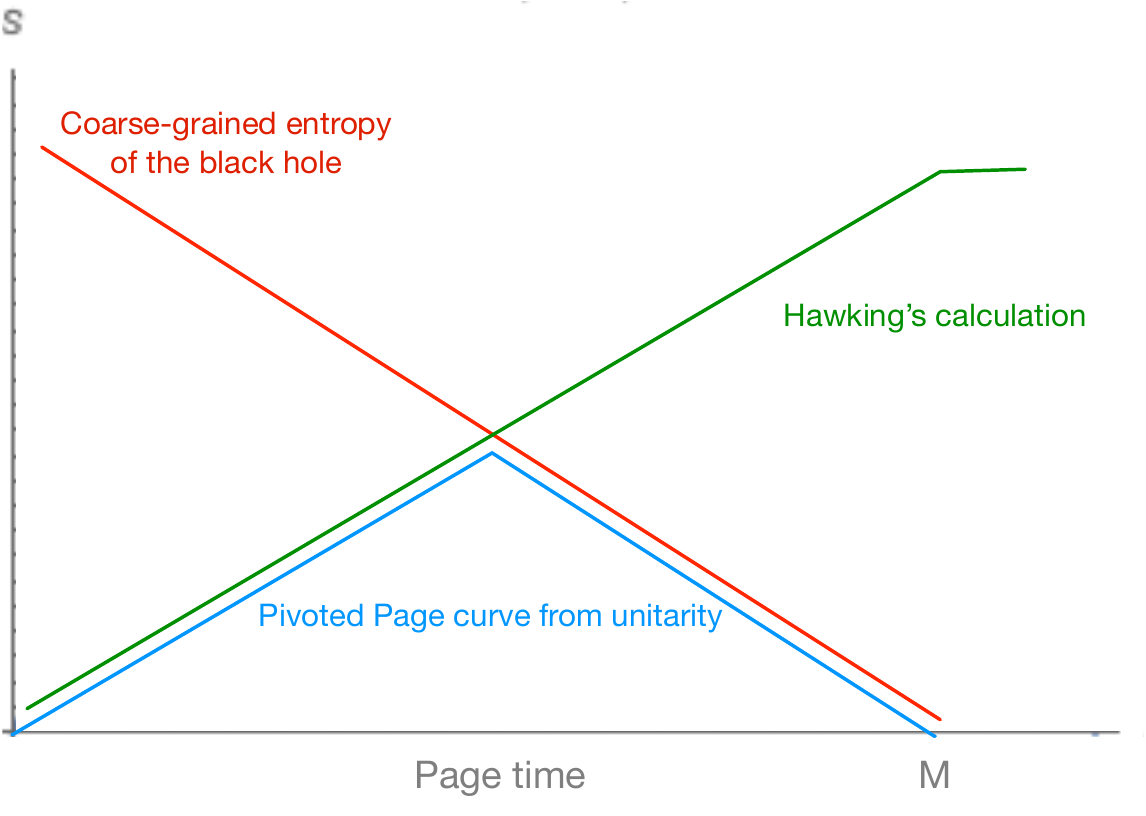}
  \caption{Schematic Page curve behaviors are shown here regarding the radiation energy from 0 to M. There are different entropy curves (including the trivial Page curve $\delta S=0$: the straight black line along the M axis). Could the entropic variation represent some direction of natural tendency?}
  \label{pagecur}
\end{figure}

Regarding black hole and its radiation in the quantum mechanism model, Page suggested in \cite{Page:1993wv} the following property of unitarity 
\bea
\exp{S_{vn}}\leq min\{d_{A},d_{\bar A}\}
\label{Unitary}
\eea
 the dimension of the smaller subsystem constraints the entanglement entropy between Hawking radiation which is thermal and black hole, and the equality could be achieved in random model. 
Thus there should be a phase transition at Page time,  when a pure state black hole evaporating out half of its mass, and the entanglement entropy would approach its maximal. Then at late time, the Page curve should be bounded by the decreasing coarse-grained entropy of black hole. See the blue curve in Figure \ref{pagecur}, for the proposed shape of Page curve from unitarity. 

The central property that Page's argument relying on is that Hilbert space of quantum system can split bipartitely as
\bea
\H=\H_{bh} \otimes \H_{R}\,,
\eea
and one can regard radiation $R$ and black hole $bh$. By a non-gravitating bath coupling, such split could be realized within the framework. The island conjecture gives a better description for the Page curve as the phase transition of QES. 

\begin{figure}[hbt]
		\centering
  \includegraphics[width=5in]{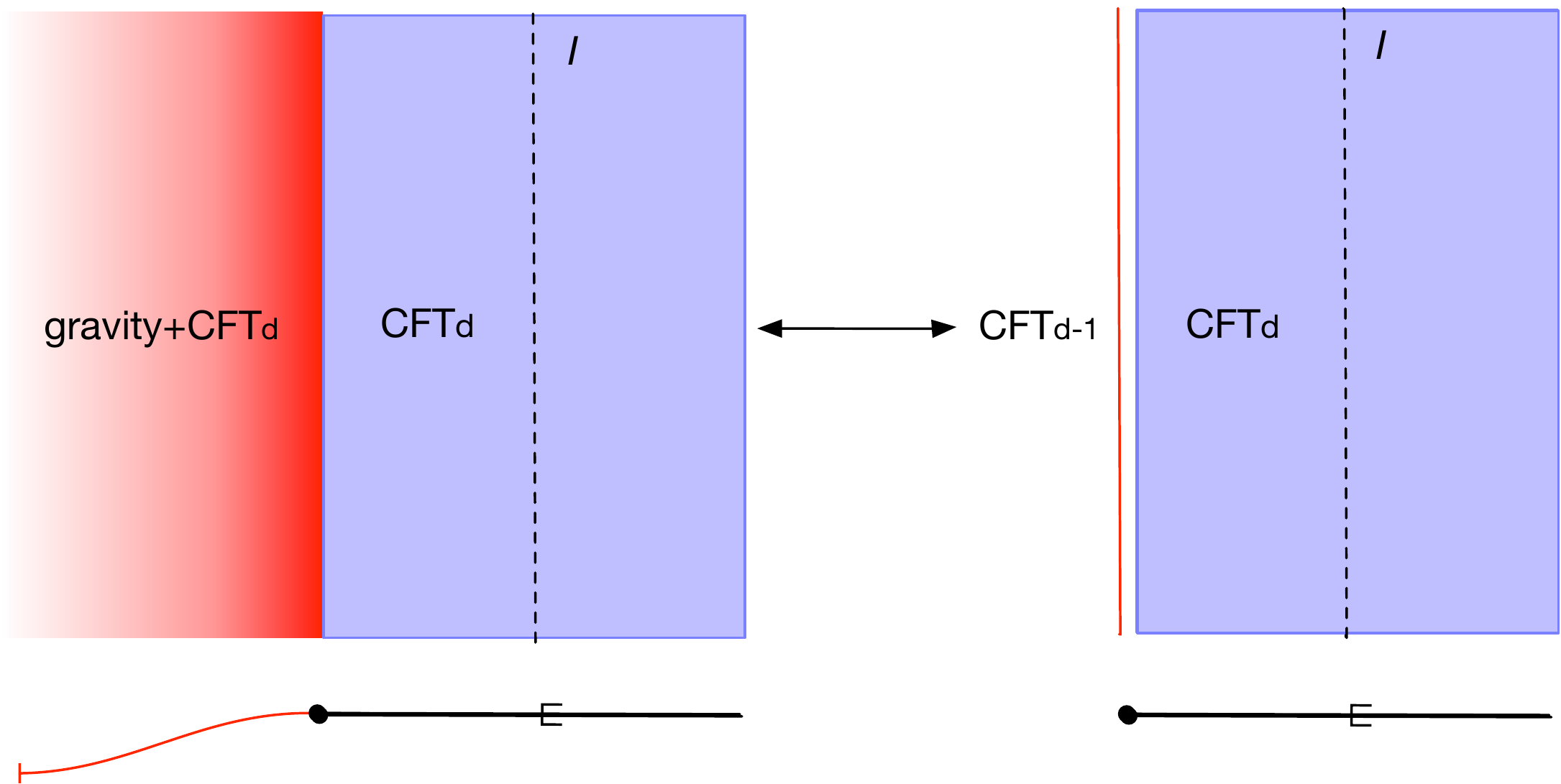}
  \caption{The typical  model of $\text{CFT}_d$ bath coupled with $\text{CFT}_{d-1}$ dual to gravity, leads to $d$-dimensional mechanical system ends at a $d-1$ dimensional defect. The dashed line``$I$" stands for imaginary interface, which split the whole system into two parts, the left part includes the defect (the black dot).}
  \label{bathdual}
\end{figure}

\paragraph{Extremal Surfaces}
In AdS/CFT, it is the geometric subject called ``extremal surface" $\chi_B$ in the bulk that corresponds to the $S_{gen}$ as
\bea
S_{gen}=\frac{A(\gamma)}{4G_N}+S_{bulk}(\gamma)\,,
\label{Sgen}
\eea
for a decomposition of boundary into subsystem $B$ and its complement $\bar B$. The classical extremal surface for static geometry is the Ryu-Takayanagi surface \cite{Ryu:2006bv} which minimizes the bulk area of $\gamma_B$, and the bulk contribution can be omitted since it is sub-leading. The HRT formula  \cite{Hubeny:2007xt} was proposed as a covariant version by adding bulk entropy as correction. 

While in quantum level, FLM was proposed in \cite{Faulkner:2013ana}, and then the Quantum Extremal Surface (QES) $\chi$ \cite{Engelhardt:2014gca}
\bea
S_{vn}= \min \left\{ext_{\chi}  S_{gen}(\chi)\right\}\,,
\label{QES}
\eea
can be applied to calculate the entanglement entropy when coupled to a bath $R$, with an extra maximin procedure to choose the one extremal surface that gives the minimum generalized entropy.
The same unitary property (\ref{Unitary}) is also the reason for the max-min procedure.

A simple special case we will use, is that for a two-sided eternal AdS black hole, the bifurcate horizon is just the extremal HRT surface $\gamma$ (which is very close to QES $\chi$) for the entanglement  between two copies of CFTs, where the holographic entanglement entropy is also the coarse-grained Bekenstein-Hawking entropy
\bea
S_{vn}=S_{BH}=\frac{A(\gamma)}{4G}
\eea
to the first order without quantum correction. This equality happens because the spacetime is stationary. That is also the situation to our Emergent Gravity theory.

For the eternal black hole case, one could maximally extend it into a two-sided spacetime.  A global pure state, the thermofield-double (TFD) state is proposed 
\bea
\psi_\text{TFD}=\sum_i e^{-\beta E^i/2}|i\rangle_L|i\rangle_R
\eea
 to be dual to such global geometry of` two-sided eternal black hole \cite{Maldacena:2001kr}.  After tracing over one side, one gets the thermal Gibbs state which is the Hartle-Hawking state, and the density matrix is
\bea
\rho_{HH}=\frac{e^{-\beta H}}{Z}\,,
\eea
with extremal surface coinciding the horizon. 
 The Euclidean path integral of eternal black hole $\rho_{HH}$ can be represented as a strip $\M_{d-1} \times \mathcal I$, with space $\M_{d-1}$ and interval $\mathcal I$ of length $\beta/2$, so the double of two strips gives the $\psi_\text{TFD}$.

\paragraph{Varying QES with bath coupling}

The realization of the Page's argument is recent through a setup of non-gravitating bath coupled to the AdS black hole. 
A relevant revisiting of the bath coupling could be find in \cite{Raju:2020smc} as in Figure \ref{bathdual}.

One can calculate the entanglement entropy of the radiation $S{(R)}$ in the bath from (\ref{QES}).
The varying of QES of an evaporating AdS black hole which is initially a pure state was considered in \cite{Penington:2019npb}. When the radiation evaporates into  a large enough auxiliary Markovian reservoir as the bath, to confirm a phase transition of QES behavior:
\begin{itemize}
  \item Early time: the QES $\chi$ is the trivial one, so the entanglement wedge of the boundary $\text{CFT}_{d-1}$ is the entire gravitating region. $S(R)$ increases with the $S_{bulk}(R)$.
  \item Late time: the entanglement wedge covers one isolated island region $I$ in interior of blackhole, and $\chi$ get transited to the the one closed to the horizon as extremal surface $gamma$ which is the boundary $\partial I$. As the black hole losses its energy, $S(R)$ decreases with the horizon area $\delta A(\gamma)$, to give an entropic gradient during evaporation. Now the black hole has become the maximal mixed state.
\end{itemize}

The important island region  for entanglement wedge of the radiation $R$
 appears after Page time \cite{Almheiri:2019hni}, through the phase transition of QES (\ref{QES}) to be
\begin{align} \label{eq:islandprescription}
S(R) = \min \left\{\underset{I}{\text{ext}} \left[\frac{\text{Area}(\partial I)}{4 G_N} + S_\text{bulk}(I \cup R)\right]\right\}.
\end{align}
 Such phase transition the quantum extremal surface requires the matter entropy should give large quantum correction of order $O(1/G)$ as first  confirmed in \cite{Almheiri:2019psf}.

\subsection{The Split Problem}
Page's argument together with the bath setup, means losing energy and information transformation from gravitating system out into some external complementary,  highly relies on the split feature of quantum system. For relativistic QFTs, 
consider the bounded region $R$ surrounded by a collar region $\epsilon$ and its complementary $R\cup\epsilon$ with the collar denoted as $\bar R_\epsilon$, all lying on on a Cauchy slice in a non-compact spacetime.  In a lattice regularization, we can decompose the total Hilbert space into $\H_R \otimes \H_{\bar{R}} $.
The total algebra could also split according to regions, such that from the algebra $ \mathcal{A}(R)$ is still ignorant of information about the other part $\bar R_\epsilon$.


We denoted the causal domain of any region $A$ as $D(A)$. While choosing another spatial region $A'$ which shared the same causal domain $D(A')=D(A)$, the entanglement entropy 
stays the same 
\bea
S(\rho_{A'})= S(\rho_{A})\,,
\eea
And under any unitary transformations $U$,
\bea
\rho_{A}'=U^\dagger \rho_A U\,.
\eea
fine-grained entropy of subsystem doesn't change either, especially under unitary time evolution.

However, when we are treating the whole system as gravitating, something could be fundamentally different, especially when two regions are along the holographic direction.

As Raju's insight \cite{Raju:2021lwh}, the split property of non-gravitational quantum field theory {could not be applied to quantum gravity.
}
Otherwise, the price is
\begin{itemize}
  \item holography is lack in the non-gravitational region
  \item breaking the gravitational Gauss law so as to the energy conservation in gravitational system but the total energy is conserved
  \item leading to massive graviton spectrum in dynamic gravity theory
\end{itemize}
 Here, we will add one new perspective here in the paper
\begin{itemize}
	
\item the bath could also supply the possible tendency violation to the original gravitating system

\end{itemize}
Inspired from this, we could reinterpret the similarity of the bath to the Maxwell's Demon to re-explain the above price of the bath setup.
\subsection{Similarity to the Maxwell's Demon}
The Generalized 2nd Law \cite{Bekenstein:1974ax} states that
the generalized entropy never decreases in an isolated gravitating system
\bea
\delta S_{gen}\geq 0\,,
\eea
showing the coarse-graining tendency for entropy to become maximal, a tendency for the close system to get equilibrium. The geometric meaning of such tendency in gravity was studied in such as \cite{Engelhardt:2018kcs, Engelhardt:2017aux}.
 However,  it's not that straightforward to apply this Generalized 2nd law to distinguish the direction of gravitational tendency such as gravitational attraction of free-falling. It can't show the tendency for open system such like coupled to external bath. Naively applying it to analyze the evaporation processes for example, one finds the shape of Page curve seems to be a contradictory.

We offer an alternative reconciliation  by awaking Maxwell's Demon, for clues to this fundamental contradiction. By our development of emergent gravity theory, we would have tools to distinguish gravitational tendency.

What important are the natural properties relevant to  about the Maxwell's Demon thought experiment. We extract them as following
\begin{enumerate}
  \item The demon can detect microscopic information and manipulate the evolving direction. 

  \item The ordinary tendency of subsystem is violated and the whole system could evolve into direction seems to be against the second law.
  \item The demon's hidden memory is always a part of the whole system for unitarity.
   \item When including back the demon, 2nd law is not actually violated.
  \end{enumerate}
 Later we will compare those similarity to  the bath setup for the Page curve.
  

 \subsection{Review: Entropic Mechanism for Emergent Inertial Force}

To explore an entanglement story for gravitational force through detailed calculation, our early work \cite{An:2018hyt} brought in the toy model in \cite{Marolf:2003sq, Marolf:2004et, Casini:2008cr}, the single-mode entangled oscillator reduced in the subsystem, to consider when the black hole background geometry thermalizes the oscillators in the local Hawking temperature, in replacement of the  common role of the Unruh temperature in common local approaches of Entropic Gravity theories such as in \cite{Jacobson:1995ab, Verlinde:2010hp}.

 
The perturbing of test particle to the nearby non-inertial trajectory in the global causal wedge by cancelling the  gravitational redshift effect, is equivalent to mountain the frequency $\omega$ fixed for the single-mode oscillator model. At the same time, the existence of the external force balanced with the gravitational force to $\bold{F}_{ex}+\bold{F}_{g}=0$, could result in a local temperature gradient $\nabla T$ measured by accelerating observers.

Using the Entanglement 1st Law, we derive the semi-classical expressions for a thermodynamic force $F_\mu$ (\ref{Fg}). We consider the specific situation when the relative entropy vanishes, which means the saturation of the Casini-Bekenstein bound (\ref{CBbound}) as in \cite{Casini:2008cr}.  The expression (\ref{Fg}) can reproduce the same covariant expression for the gravitational force $\bold{F}_{g}$ in GR as tested in the spacetime of asymptotic flat Schwarzschild black holes and Newton's 2nd Law $F=ma$ for Rindler space. This mechanism works generally beyond near horizon region, and it also turns the derivation of Einstein equation from null screen \cite{Jacobson:1995ab} to the time-like screen \cite{Verlinde:2010hp} valid, as we later proved in \cite{An:2020ncr}. 
 
Therefore,  the resulting temperature gradient $\nabla T$ for the local observer Alice is the reason for the thermodynamic force $ F_{\mu}$ as well as an entropic gradient $\nabla S$, in the spirit of the Onsager reciprocity relations. Or to the view an asymptotic observer Bob,  the temperature is fixed to be $T_H$ but the ADM energy gets perturbed during the process such that the isoenergic process for Alice becomes isothermal process for Bob.  


\paragraph{New approach through the Entanglement 1st Law}
What really matters here are two following points: 

(i) We develop a modular Hamiltonian approach to extract gravity as an extra work term that we seek from the entanglement first law.
The entanglement first law states that if $\rho_R(\lambda)$ of a state in the subsystem $R$ varying with one parameter $\lambda$, to the first order perturbation $d\lambda$ at $\lambda=\lambda_0$, we always have the following equation
\bea
\frac{dS(\rho_R)}{d\lambda } =\Tr \left(\frac{d\rho_R}{d\lambda} {K_0}\right)
\label{firstlawK}
\eea
or we can rewrite it as
\bea
dS=d\braket{K_0}
\eea
where $K_0=-\log{\rho_R(\lambda=\lambda_0)}$ is the modular Hamiltonian of the initial state. A detail proof can be find in \cite{VanRaamsdonk:2016exw}. 

As a consequence of (\ref{firstlawK}), we could take the parameters such as temperature $T$ in $K_0=H/T$ out of the derivative
\bea
TdS=d\braket H\,.
\eea
 during certain processes. We show the gravitational force can be reproduced in an entropic mechanism
\bea
F_{\mu}=T\nabla_\mu \braket{K_0-K_1}_1\,.
\label{Fg}
\eea
as the contribution exactly from the difference between the excited modular Hamiltonian of the form $K_1=K_0+O$ in \cite{Balakrishnan:2020lbp,Arias:2020qpg} and $K_0$ is the vacuum state, where $T=\frac{\kappa}{2\pi V(r)}$ represents the local temperature with the surface gravity $\kappa$ and the redshift factor $V(r)$.

The reason to base on the entanglement first law is that modular Hamiltonian of the Hartle-Hawking state reduced in external of horizon
\bea
\rho_{HH}=\frac{e^{-\beta H}}{Z}
\eea
is simply $K_0=\beta H=H/T_H$, which is exactly the operator for the time Killing vector $\partial_t$.

(ii)We show the gravitational force is not entropic by keeping entropy and thermal distribution invariant during free-falling, until an isoenergic process happened to overcome the gravitational redshift effect. 

The Casini's version of the Bekenstein \cite{Casini:2008cr} bound is
\bea
 \Delta S\leq \frac{\Delta \braket{H}}{T_H}=\frac{\Delta \braket{H_A}}{T}=\lim_{upper}\Delta S\,,
\label{CBbound}
\eea where $\Delta S=S_1-S_0$ and $\Delta \braket{H_A}=\braket{H_A}_1-\braket{H_A}_0$ for the local $H_A=H/V(r)$ and local temperature $T=T_H/V(r)$ are the renormalized entanglement entropy and the renormalized energy respectively of excited state $\rho^1_R$ from the vacuum $\rho^0_R$. When this bound is saturated, (\ref{Fg}) leads to the thermodynamic expression

\bea
F_\mu=\lim_{upper}\Delta S\nabla_\mu T= \frac{\nabla_\mu T}{T}\Delta \braket{H}\,,
\label{uniexforce}
\eea
for $F_{\mu}$ to compare with the local inertial force $\textbf{F}_{g}=-ma_\mu$ and the entropic force formula, where the renormalized energy $\Delta \braket{H}$ is statistics-dependent with the statistical factor $\frac{1}{1\pm e^{-\omega/T}}$.
After adopt $T=\frac{T_{H}}{V(r)}$ with $V(r)=e^{\phi(r)}$ being the redshift factor to the generalized Newton's potential $\phi$, this expression is exactly the local gravitational inertial force.

\paragraph{Entropic direction}

When the local Hawking temperature is very low such that the distribution factor $e^{-\omega/T}\ll1$, $\Delta\braket{H_A}$ for the co-moving observer Alice's local measurement $H_A$ stays almost the same when Alice moves between nearby local static trajectories labeled by $\lambda'$ and $\lambda$ in such frequency-fixed process:
\bea
\Delta\braket{H_{A'}}\approx \Delta\braket{H_A}
\eea
so the gravitational redshift effect to the frequency is canceled.
And the entropy bound varies in this the isoenergic process approximates to
\bea
d\Delta S\approx\frac{\Delta\braket {H}}{T_H}d V(r)
\eea
Thus this process approximates to the isoenergic process in low temperature limit. So (\ref{uniexforce}) leads to a minus sign in 
\bea
dW_g\approx-T_Hd\Delta S=-dQ\,
\eea
which means the opposite direction to the variation of the entropy bound $\Delta S$, while $F_{ex}$ is along the direction of the maximal heat flux $dQ=T_Hd\Delta S$ into the bath during such processes.

\section{Coarse-graining Arrow and the 2nd Law Tendency}
\label{coarse}

Above, we actually pray for an alternative method to distinguish the direction of gravitational tendency.

Here, an local external force
\bea
\bold F_{ex}=T \nabla_{\mu} S
\eea
is required to equilibrate the emergent gravitational inertial force $\bold F_g$.  At the same time, the entropic gradient $\nabla_{\mu} S$ is also emerged due to the energy flux $dE$ as well as the maximal heat flux $d Q=T_Hd\Delta S$ into the external auxiliary system.

\subsection{Euclidean state: Fixed Thin shell attracted by a black hole}

\begin{figure}[hbt]
		\centering
  \includegraphics[width=3in]{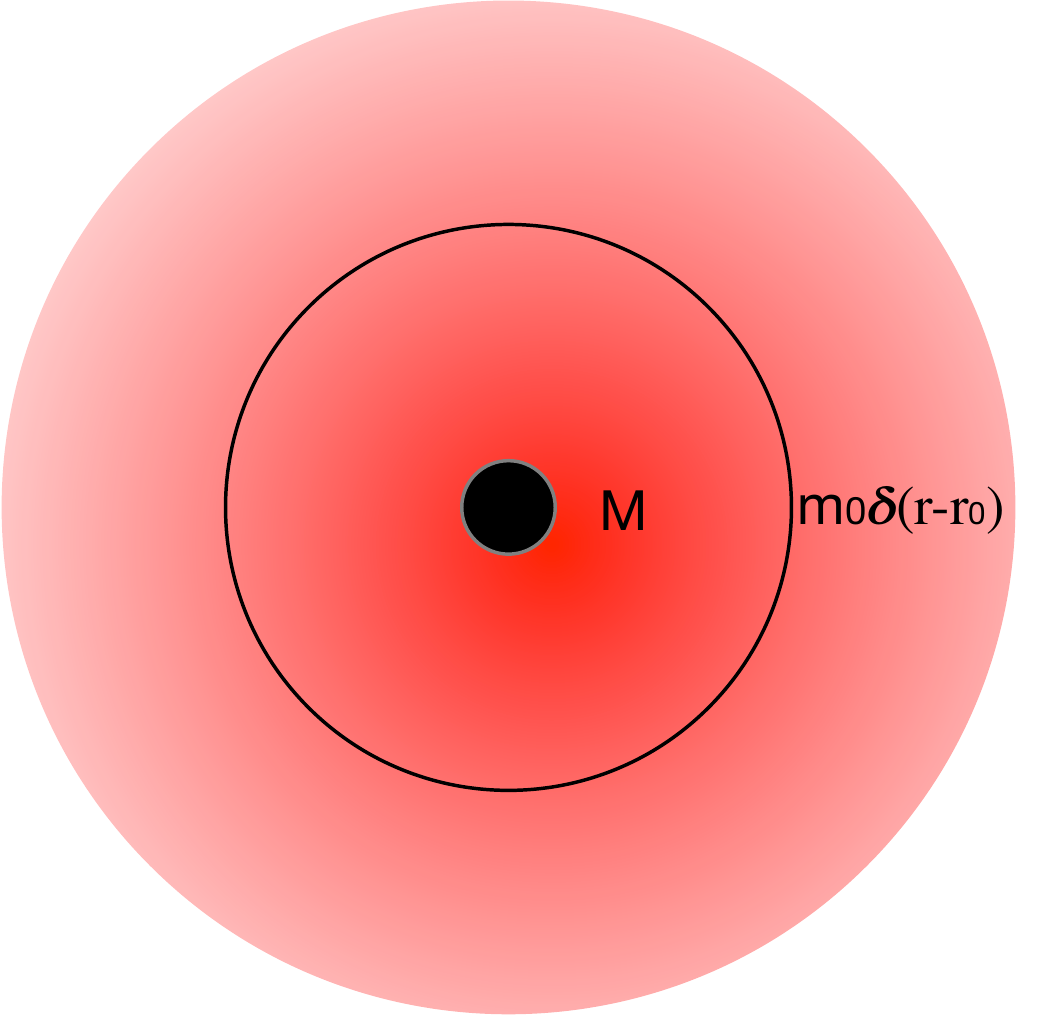}
  \caption{The static picture of a spherical black hole $M$ attracting a test spherical thin shell of matter $m_0$ fixed at $r_0$. In a local inertial frame, $m_0$ is accelerating. In Euclidean picture, we view such state as infinitesimal perturbation thermalized with a local Hawking temperature.}
  \label{m0}
\end{figure}

To give a straightforward image of the simple situation we re-investigate, we just consider a spherical massive thin layer of matter fixed at radial $r_0$ outside an eternal black hole horizon as shown in Figure \ref{m0}. 
Our picture is quite similar to the consideration of adding matter outside of horizon as perturbation \cite{Chandra:2022fwi} of pure EOW(end-of-world brane) state solutions \cite{Cooper:2018cmb}, which is based on the collapsing thin shell black hole solutions for pressure-less fluid  studied in \cite{Keranen:2015fqa,Anous:2016kss}. While they add matter following geodesic in pure state, we fix the matter static outside and take the maximal mixed state whose entropy is up to the bound.

This static picture could be taken as Euclidean state, where everything in space  doesn't not evolve with the real time $t$.
To calculate the entropy, a short cut is that, we are able to base on the perturbation of the Euclidean black hole state. The perturbation  on this Hartle-Hawking state is
\bea
\delta S_{bh}=\beta_H m\,,
\eea
where $m$ is the ADM mass of local measured energy $m_0$ and $\beta_H=1/T_{H}$ is the inverse of Hawking temperature, from the 1st law of black hole thermodynamics $dM=TdS_{bh}$, where 
we take $\delta M\rightarrow m$.

Since we know  a Euclidean path integral, will give the Bekenstein-Hawking entropy formula 
\bea
S_{BH}=\frac{A}{4G}\,.
\eea
where $A$ is the area of horizon $\Sigma_{r_s}$, thus there is a hidden equality holds:
\bea
\beta_H m =\frac{\delta A}{4G}\,,
\label{m=dA}
\eea
which is exactly the saturation of the Casini-Bekenstein bound (\ref{CBbound}) for $\Delta \braket H\rightarrow m$. We have check the equality (\ref{m=dA}) for asymptotic flat Schwarzschild black hole in \cite{An:2020ncr}.

 The new application here is, the 1st law of black hole thermodynamics not only about two black hole states. We explain this amount of outer entropy as the infinitesimal perturbation outside a black hole horizon
\bea
S(M,m)=\frac{\delta A(\Sigma_{r_s})}{4G} \propto GMm
\eea
to be associated with gravitational. attraction. This amount of outer entropy when equilibrating is just same amount of a large black hole.

 The difference to the thin shell pure state is that our picture is not stable if the thin layer of matter is not fixed, because the spherical layer gets attracted of the black hole: therefore it requires an external force to maintain equilibrated against the local inertial force.

\paragraph{Connection to the thin shell pure state}

Here, we simply refer to the results of coarse-graining a pure state in 
\cite{Chandra:2022fwi} to compare with our setup of perturbation on maximal entanglement states.    

When adding matter outside horizon to the thin shell state as $\rho$,  the coarse-grain map $\mathcal C$ will give a correction of total coarse-grained entropy for outer coarse-grained entropy 
\bea
\delta S_\text{coarse}(\rho)=\frac{\delta A(\mu_a)}{4G}\,.
\eea
 as the variation of apparent horizon $\mu_a$.  It is just the correction to the horizon $\Sigma_{r_s}$.
 
One can do block-diagonal projection to get a state  $\bar\rho$ form $\rho$. The von Neumann entropy of the state $\bar\rho$ is just
 \bea
 S_{vn}(\bar{\rho})=\frac{A(\gamma')}{4G}
 \eea
  which is sub-dominant extremal surface for a pure state $\rho$ but dominant for $\bar \rho$. While, here we would tell the Casini-Bekenstein bound of entanglement entropy is the same amount of outer entropy.

And $\mu_a$ becomes the real event horizon after the thin layer is absorbed into the horizon, a way of coarse-graining with external system, one can transform 
\bea
\mu_a\rightarrow\Sigma_{r_s}'\approx\gamma'
\eea
With such a bath or with external force for equilibrating, one can really transform the 
\bea
S_\text{coarse}\mapsto S_{vn}
\eea
to get a maximal entangled-state to be $\bar \rho$, with the maximal outer entropy
\bea
\delta S_\text{coarse}\mapsto \lim_{upper}\Delta_{} S
\eea
transform into the entanglement entropy bound as we point here.

That's the reason why so far we only need to rely on the special static situations, like two-sided black hole which is in thermal equilibrium, such that entanglement entropy is equal to the coarse-grained entropy. It was formed for perturbing the two-side black hole, such that holographic entanglement entropy is also the thermal entropy
\bea
S(M,m)\equiv \lim_{upper}\Delta_{} S=\delta S_\text{coarse}=\frac{\delta A(\mu_a)}{4G}\,.
\label{Sentcoar}
\eea
And the entanglement entropy bound is saturating to the infinitesimal perturbation as checked in \cite{Blanco:2013joa}. From here, we only use the symbol $\Delta_{} S$ to represent the entanglement entropy bound
 since we only work with the saturation.  


\subsection{Entropy Gradient associated with gravitational attraction}

In general, we would suggest (\ref{Sentcoar}) is the amount of entropy associated with two subjects attracting each other. To the case very small object with local measured mass $m_0$ with the gravitational potential $\phi$ is attracted by large black hole $M$, we propose the entropy associated with such attraction is:

\bea
S(M,m_0 e^{\phi})={m_0 e^{\phi} \over T_H}=8\pi G M m_0 e^{\phi}
\eea
 to the leading order, where $m=m_0 e^{\phi}$. Taking $m_0e^\phi$ as infinitesimal perturbation, such entropic bound is indeed saturated relying on \cite{Blanco:2013joa} and it is covariant. To modify it in isoenergic processes as shown in \cite{An:2020ncr} will simply lead to the correction to the Verlinde's proposal of entropic gradient \cite{Verlinde:2010hp} in generic situation to be
\bea
\nabla_\mu S&=&\frac{1}{T_H}m_0e^\phi\nabla_\mu\phi\,.
\label{nSphi}
\eea
 The generalized gravitational potential is defined from the time Killing vector $X_\mu$:
$\phi\equiv \frac{1}{2}\log\{-X_\mu X^\mu\}$ and we rewrite the redshift factor as $V=e^\phi=\sqrt{-g_{tt}}$. Therefore we can derive local gravitational force $\bold F_{g}=-\bold F_{ex}$ indirectly 
from the entropic force formula
\bea
\bold F_{ex}=T\nabla_\mu S
\eea
by using local Hawking temperature $T=T_{H}e^{-\phi}$, which eventually gives consistent results with textbook derivation of gravitational force in GR (see reference textbook \cite{Carroll:2004}). As well, one can use Unruh temperature, to derive $F=ma$ in Rindler space. Such amount of entropy updates the derivation of Einstein equations from thermodynamics beyond horizon region. Those results can be found in our early paper \cite{An:2020ncr}.

The entropy gradient shows the direction of gravitational potential, as the falling tendency in spacetime. 
It will not be hard to apply this entropic mechanism to the AdS-Schwarzschild black hole of the same form of metric (set $l_{AdS}=1$ for convenient)
\bea
ds^2=-f(r)dt^2+{1\over f(r)}dr^2+{r^2}d\Omega_{d-2}
\eea
where
\bea
f(r)=1+r^2-\frac{16\pi GM}{(d-1))\Omega_{d-1}}\frac{1}{r^{d-2}}
\eea
and $M$ is the ADM mass obeying the 1st law of black hole thermodynamics, with the redshift factor
\bea
V(r)=\sqrt{f(r)}
\eea 
increasing to infinity (in asymptotic AdS) rather than 1 (in asymptotic flat spacetime) as $r\rightarrow \infty$ near the boundary, which stands for the infinite potential barrier.

\subsection{Distinguish the Coarse-Graining Arrow}

The center problem is how we can distinguish falling tendency from entropy variation. After that, we may relate this mechanism to general gravitational tendency such as Hawking radiation in next section.

With the equilibrating by external system, one could evolve Euclidean states in quasi-static processes like reversible heat machine. Then, we will orient the variation of the entropy bound in the direction of increasing heat flux.
In the spirit of the Onsager Reciprocity Relations, we would say $\bold F_{ex}$ reveals the invisible  $\bold F_g$ for the geodesic, as well as those thermodynamic gradients.


\begin{figure}[hbt]
		\centering
  \includegraphics[width=5in]{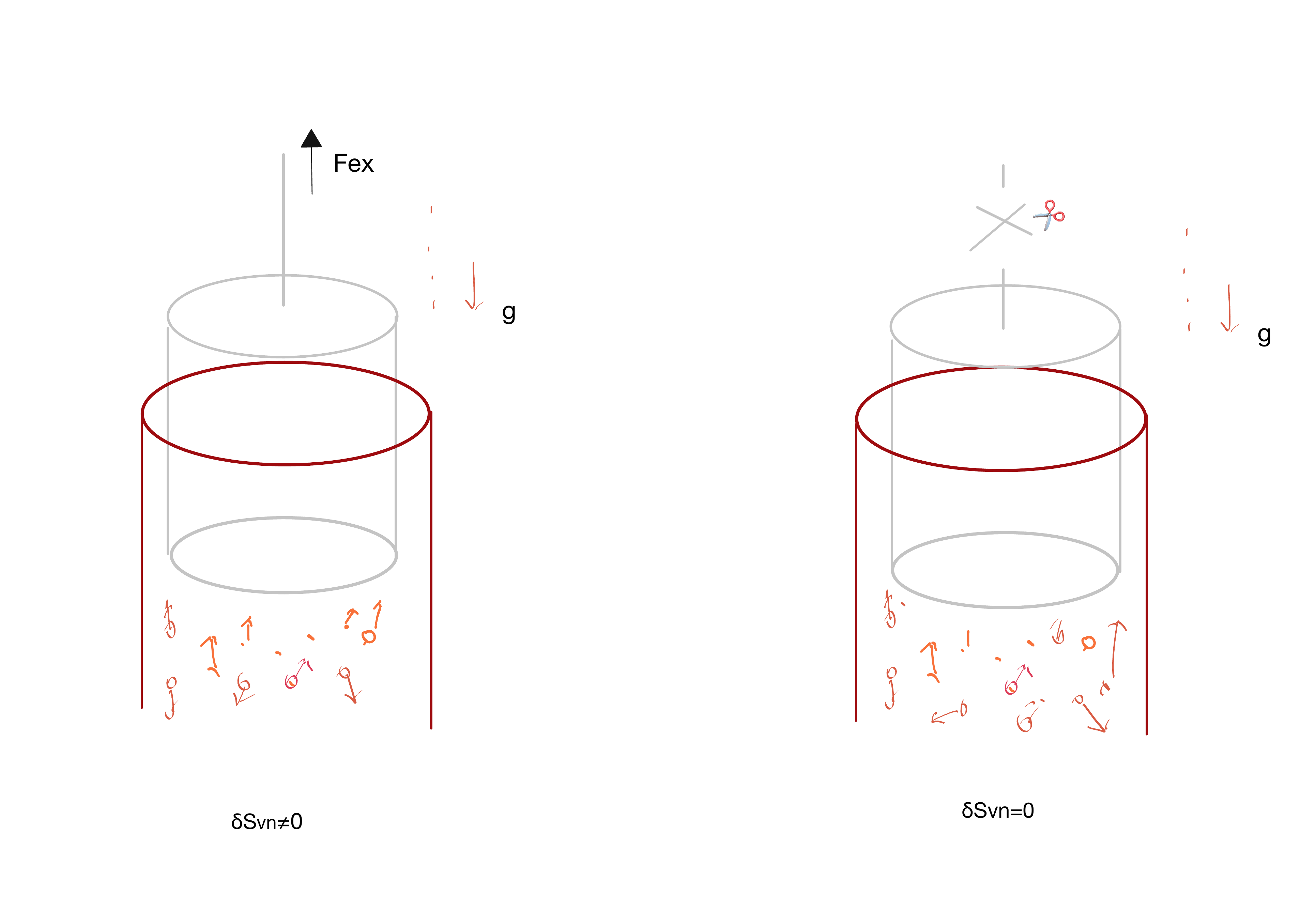}
  \caption{
 In an isolated system without external force by the top string (Right), gravitational evolving such as free falling doesn't change fine-grained entropy even some of gravitational potential gets dissipated thermally. While, entanglement entropy can be changed with transforming energy into external system (Left). 
 One way to tell falling tendency through entropy variation, is to bring in the presence of external force for equilibrating, and then take the split from the external system. 
  }
  \label{engines}
\end{figure}

The fine-grained entropy doesn't change in an isolated system.  Whereas,  when switching the adiabatic to the quasi-static, the variation of fine-gained entropy could be used to indicate the coarse-graining direction. Our target is to give an analogue to show the direction of gravitational tendency is actually 
\bea
F_{g}\propto-\delta_{\lambda} S(M,m)\,,
\eea
noted here it is also along the opposite direction of holographic entanglement entropy variation $\delta S_{vn}$ since the equality (\ref{Sentcoar}). And we note here that, the same negative entropy gradient happens at late time Page curve, so could
\bea
F_{g}>0
\eea
is repulsive? We will discuss in the next section.

\paragraph{Analogy to Heat Engines}

To make an analogy to a thermodynamic situation, we can show this gravitational falling situation as a reversible ideal heat engine in Figure \ref{engines}. 

For a closed system, there no fine-grained entropy changes in adiabatic process, but the coarse-graining entropy gets maximized
 for the 2nd law tendency
\bea
\delta S_{vn}=0 \text{ but } \delta S_{\text{coarse}}>0\,.
\eea
\paragraph{Distinguish Coarse-graining direction of Adiabatic from Quasi-static} 
In thermodynamics, one way to tell the evolving direction of adiabatic process is to switch the system quasi-static.

At some time before equilibrium, one can freeze the evolution of system by adding a piston with external force $F_{ex}$ for equilibrating and split the whole system bipartitely into $A$ and $\bar A$, so one have
\bea
F_{{A}-\bar{A}}+F_{ex}=0
\eea
where $F_{{A}-\bar{A}}$ in the middle is emergent from such as the density gradient between two subsystems.

\begin{figure}[hbt]
		\centering
  \includegraphics[width=5in]{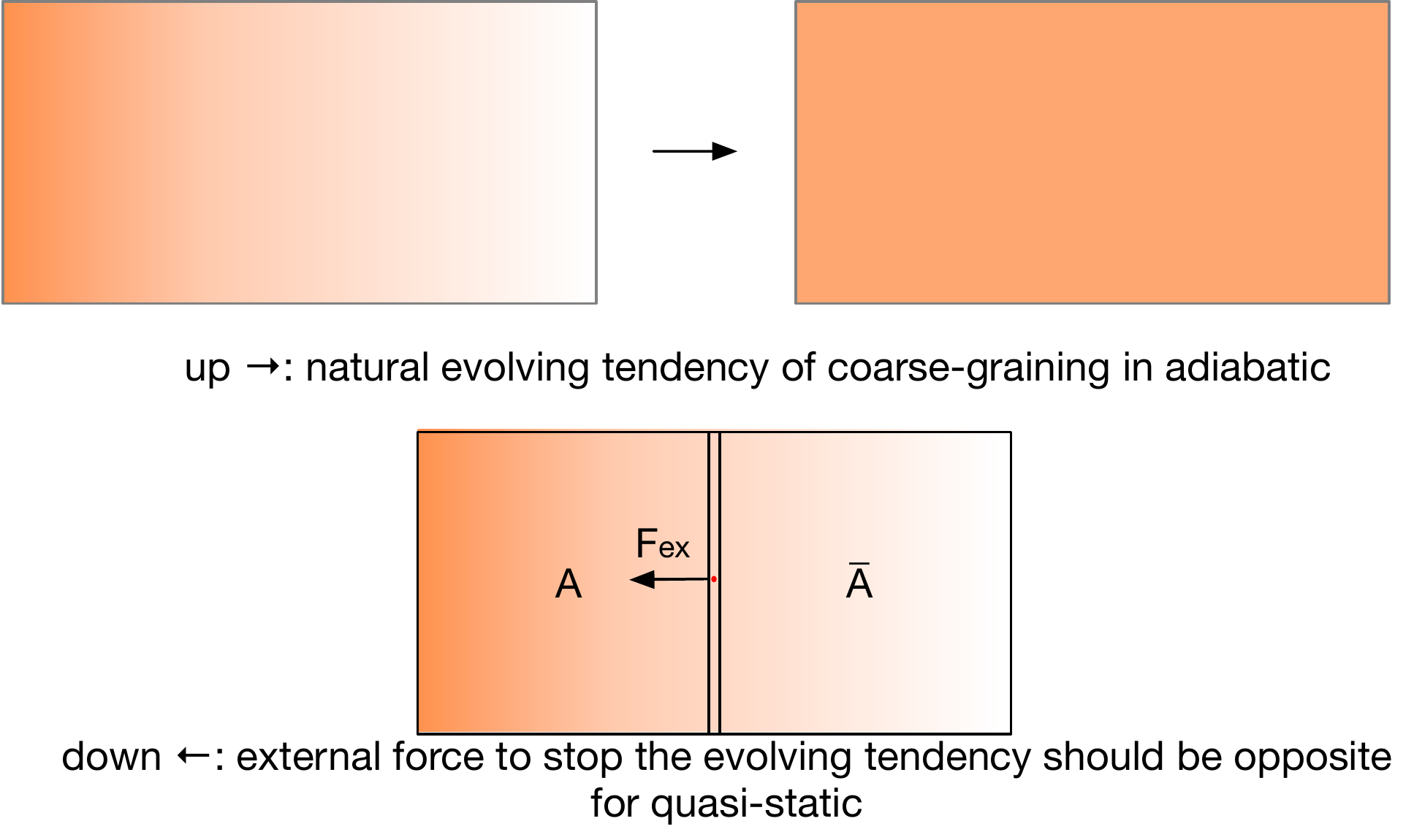}
  \caption{We refer to a simple fact on how $ F_{ex}$ to maintain quasi-static can indicate the evolving direction of adiabatic process (which could be close to equilibrium). From the view of Onsager reciprocity relations for dissipating processes out of equilibrium, the density gradient/temperature gradient gives thermodynamic force, which can be stop by an external force. After we equilibrate the isolated system bipartitely with $F_{ex}$ to stop the evolving, $F_{ex}$ ($\leftarrow $) is always in the opposite direction of the evolving direction ($\rightarrow $). In the same way to gravity, if we stop the free-falling for quasi-static, we are able to tell the falling tendency from opposite direction of  $\bold F_{ex}$.}
  \label{Evolve}
\end{figure}

The original tendency could be distinguished as the opposite direction of the external force, as we show in Figure \ref{Evolve}. After the split, $\bar A$ plays the same role of the bath, which includes all the environment as well as  $F_{ex}$ in $\bar A$ to evolve subsystem $A$.


Switching back to the adiabatic process, once we release the piston freely, where the coarse-graining take place to maximize the entropy. And one can use $F_{ex}$ shows this coarse-graining tendency $\delta S_\text{coarse}>0$.

\paragraph{Back to gravitational tendency}

The thermodynamic situation tells us the evolving tendency is opposite to the direction of $F_{ex}$. Here we show this entropic mechanism can also distinguish the free-falling tendency  which is adiabatic without $F_{ex}$, after the analogue to heat engines.  Informationally, $F_{ex}$ is something like auxiliary external system we ignore in the whole system. Without $F_{ex}$, nothing really changes QES and the entanglement entropy. 

Similarly, such external force is due to the existence of an auxiliary system outside of gravitating system.
\bea
F_{ex}\propto\delta_\lambda \Delta S
\eea
through the entropy variation.

However, because external force is doing negative work to the system $A$  when extracting  energy out $\delta \Delta \braket H<0$, the system has the tendency to evolve along the direction to decrease holographic entanglement entropy bound
\bea
\delta_\lambda \Delta S<0\,.
\eea
The analysis is done on the two-side black hole, where holographic entanglement entropy is also the thermal entropy. We see perturbation on the states gives the exact expression of gravitational attraction \cite{An:2018hyt} from our theory of Emergent Gravity.

Finally, we turn the system back to be adiabatic by withdrawing the external force, where the coarse-graining take place to maximize the entropy up to the bound. The free falling towards the black hole goes to maximize the generalized entropy
\bea
\delta S_{\text {gen}}\geq0\,,
\eea
in the Generalized 2nd law direction \cite{Bekenstein:1974ax},  while the falling direction of $F_g$ is simply along the direction of coarse-graining entropy increasing $\delta S_{\text{coarse}}>0$.

We would conclude in this easy example, the direction of $F_{ex}$ is along the same direction of entanglement entropy bound increasing?
 While, the gravitational force is in the opposite direction,   which is the coarse-graining direction if we withdraw $F_{ex}$ to switch to adiabatic process.



 \subsection{Non-variation of Extremal surfaces:  Natural Tendency of trivial Page Curve}

Without the bath, non-variation of extremal surface is indeed the feature of holographic entanglement entropy in AdS/CFT, which corresponds to the trivial Page curve  for the evaporation
\bea
\delta S_{vn}=0\,.
\eea 
Similarly, this is also the reason not able to get an entropic mechanism of free-falling.




If all the region is gravitational, one can't split the Hilbert space easily. So, we will also get a trivial Page curve when we take the bath gravitating. This could be the fundamental property of a closed QG.

By changing the boundary condition and coupling to a bath, somehow, one has changed the ordinary tendency of gravitating system. So one only gets entropy variation due to the system is open to exterior. That already changes the original tendency from $\bold F_{ex}$. The entropic variation, no matter increasing or decreasing,
\bea
\delta S_{vn}>0 \text{ or } \delta S_{vn}<0
\eea
 is violating such natural tendency.

Above, it is quite motived to introduce the concept of Maxwell's demon to analyze the Page curve pivoting.

\section{Tendency of Page Curve}

There are only two directions of entropy variation: increasing or decreasing, just like there are two directions of gravitational tendency: attractive and repulsive. Apply the above method from gravitational falling to analyze black hole evaporation, 
\bea
\delta
\Delta S\rightarrow \delta S_{vn}=\delta S(R)\,,
\eea
 we may able to deduce prediction from Page Curve: the possible turning point of gravitational tendency,  because of the appearance of the island saddle for phase transition of entanglement entropy.

\begin{figure}[hbt]
\centering
  \includegraphics[width=3in]{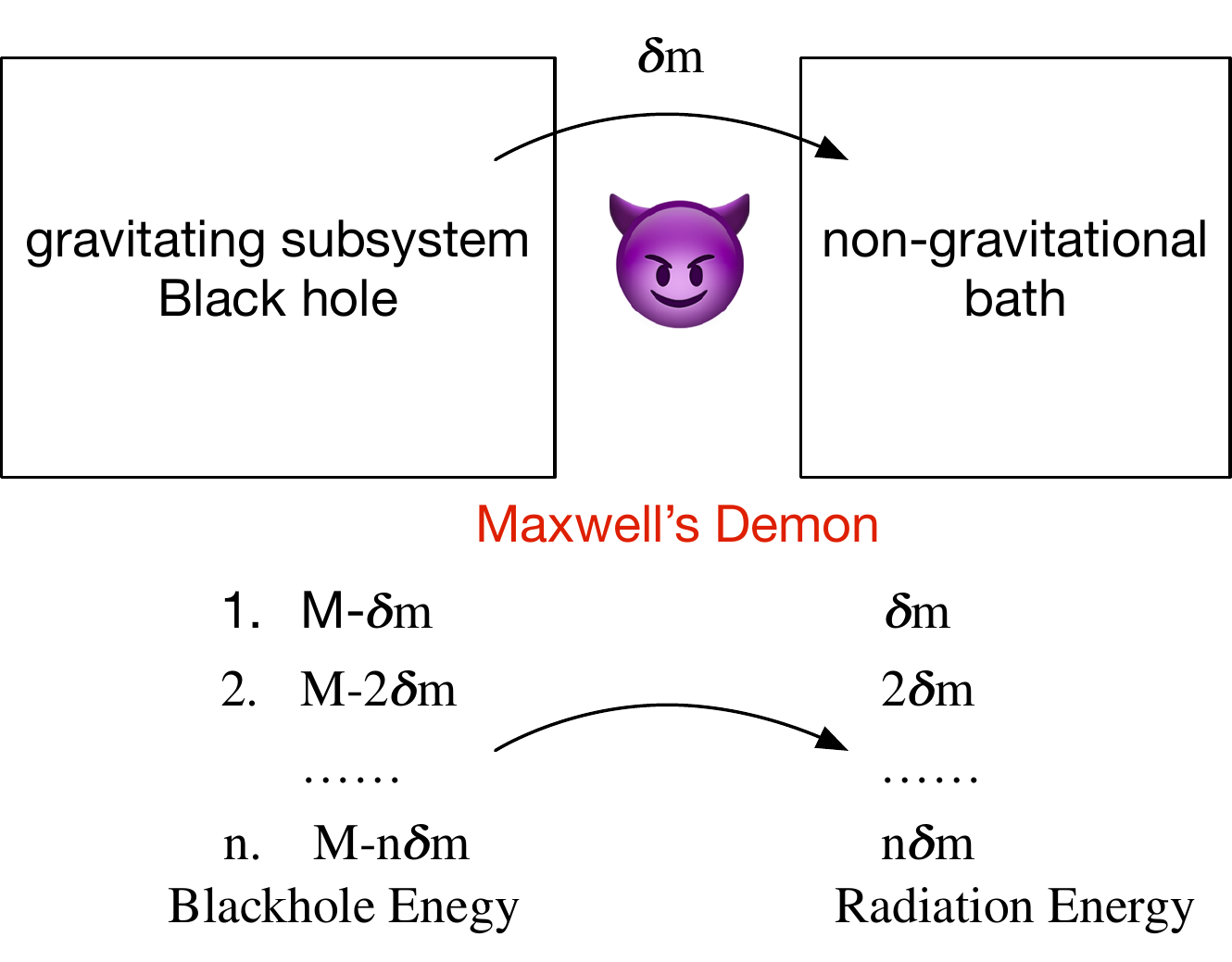}
  \caption{We consider the evaporating process when the energy is continuously extracted out into the bath iteratively in quasi-static. The question is: what is extra external influence should the demon provide? Is the energy extraction against the original tendency or follow the tendency?}
  \label{massdemon}
\end{figure}

To judge if the evolving direction is against the tendency, we consider the process when radiation energy is iteratively extracted out into the other part of system  as in Figure \ref{massdemon}. The strategy is to use the help of the demon or bath coupling, no matter if the process is against the original evolving tendency, then we try to analyze what did the demon really do.

 We will start by analyzing the late time and give an exhibition from the Dyson's sphere model.

\subsection{Late time first}
First of all, the late time period is just the same situation for our theory of emergent gravity theory for perturbation on the two-sided black holes. The black hole exterior region has been just maximal entangled. At each slice of late time, the black hole entanglement wedge is some Euclidean state, which is also very closed to the Hartle-Hawking state
\bea
\rho_{HH}=\frac{e^{-\beta H}}{Z}\,,
\eea
with extremal surface coinciding horizon, the late time is similar to our Emergent Gravity situation.

At late time, from the competition between the entropy radiation and area entropy decreasing, one need to include the island contribution in the computation of
QES, for the extremal
\begin{align} \label{eq:islandprescription}
S(R) = \min \left\{\underset{I}{\text{ext}} \left[\frac{\text{Area}(\partial I)}{4 G_N} + S_\text{bulk}(I \cup R)\right]\right\},
\end{align}
Reminded that the location of the minimal QES is closed to the horizon after the Page time, due to the homology constraint for RT surface and unitarity, $S$ is decreasing with the area of horizon shrinking. 
When S(R) decreases, we would claim at least at late time with island, we can argue there is an external force to extract radiation out of gravitational system
\bea
F_{ex}\propto\delta_\lambda S(R)\approx \delta_\lambda \frac{\text{Area}(\Sigma_{r_s
})}{4 G_N}<0
\eea
we get a negative entropic gradient at late time.

In other word, the radiation is rushed out even though the traditional gravitational attraction is inward. Thus a net force in the opposite direction: outward. Quite a surprise to see the gravitational tendency to the radiation is repulsive and outward
\bea
F_g>0\,.
\eea
Since the external force is against the direction of evaporation in the late time, we can see the tendency is obeyed by the radiation in the late time.

\subsection{An exhibition by the Dyson's Sphere}
We reuse the alternative setup that a Dyson's sphere collecting radiation in \cite{Bousso:2019ykv,Bousso:2020kmy}. The Dyson's sphere is set near the boundary of AdS. The Dyson's sphere picutre could exhibit such situation in a close QG system. 

We start from a pure state big black hole in AdS, which already achieves equilibrium  with the boundary. Ever since, the spacetime stays static. So we know the von-Neumann entropy of boundary CFT will not change. And let's suppose the state of the dual quantum CFT stays pure.

 In the light-cone coordinates
\bea
u=t-r,v=t+r
\eea
the reflecting AdS boundary condition is
\bea
T_{uu}|_\epsilon=T_{vv}|_\epsilon>0\,.
\eea
There are continuous Hawking radiation reflected back by the boundary so the ingoing mode has the same amount as outgoing modes. So the reflecting boundary condition of asymptotic AdS serves as a closed box, which is the ordinary AdS tendency.
\begin{figure}[hbt]
\centering
  \includegraphics[width=2.5in]{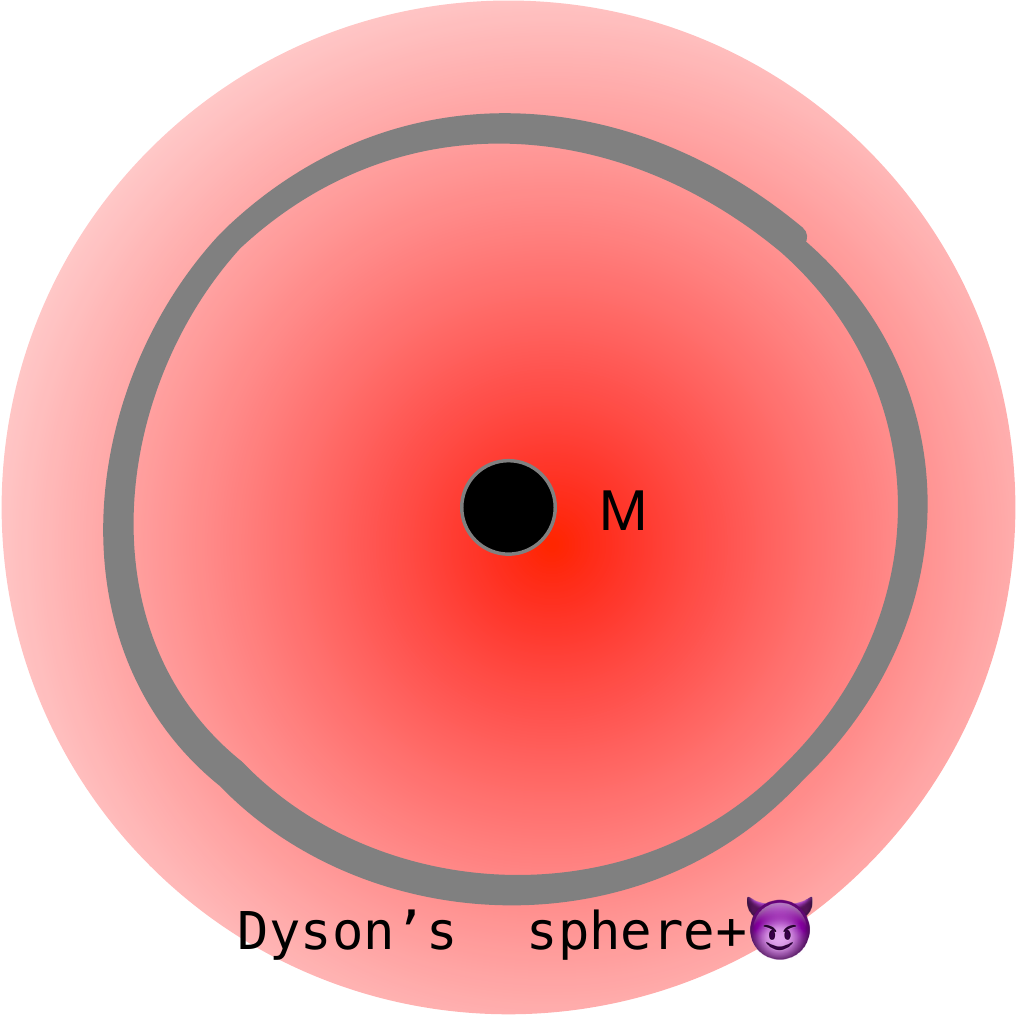}
  \caption{We use the Dyson's sphere to collect radiation near the boundary, and the demon can change the tendency of radiation like the bath to stay inside and evolves the black hole evaporating out. Differently, it means the bath is also in same bulk.}
  \label{Dyson}
\end{figure}

\paragraph{Turn on the Coupling}
At some time, when we start to turn on the function of collecting radiation, letting the radiation no longer go back to the black hole from the boundary. 

The Dyson's sphere is used to collect the Hawking radiation from inside.  Then the radiation can accumulate on the Dyson's sphere layer by layer as in Figure \ref{Dyson}. So we see the black hole energy gets lost and the horizon gradually shrinks. Macroscopically, the sphere should exert an external force $\bold F_{ex}$
 to stop the ingoing mode at the cut-off $z={\epsilon}$. And the above subsection suggests the direction of $\bold F_{ex}$ and the possible gravitational tendency change from Page curve. 

That is equivalent to say we manipulate the original tendency of AdS by a Maxwell's Demon on the sphere and it has the same effect of absorbing boundary condition if we cancel out all the ingoing mode,
\bea
T_{vv}|_{\e}=0
\eea
 such that $\lambda$ is synchronic with the boundary time $t$. And the effect of Maxwell's Demon is similar to that we turn on the coupling with external auxiliary non-gravitating baths for semi-classical approach.

\subsection{No massive graviton in Euclidean Evolution}

In short, the massive graviton argument \cite{Geng:2020qvw} is that the broken of the Gauss Law
\bea
\partial_a T^{ab}\neq 0,
\eea
leads to no massless graviton mode. 
\begin{figure}[hbt]
\centering
  \includegraphics[width=2in]{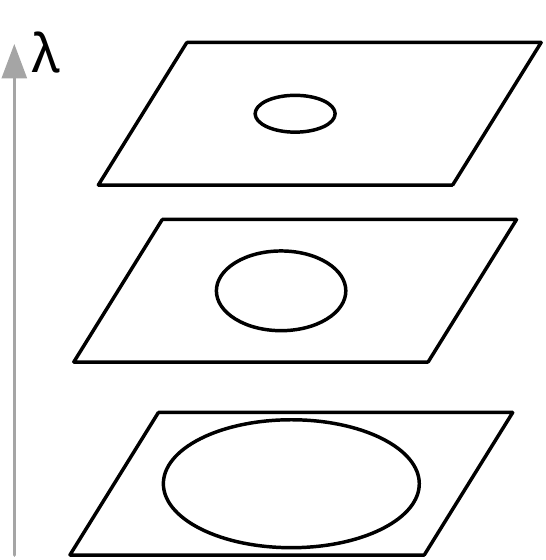}
  \caption{Instead of real time evolution, the gravitational evolution by the demon is just for effective theory of gravity, without the cutout part of UV. On each slice, Euclidean state is actually parameterized by $\lambda$ rather than real time.}
  \label{EuclEvol}
\end{figure} 
In the light-cone coordinates
\bea
u=t-r,v=t+r
\eea
the absorbing boundary condition is that
\bea
\frac{dE}{dt}=\braket {:T_{uu}:}-\braket { :T_{vv}: }
\eea
with ingoing part $T_{vv}=0$, which breaks the gravitational gauss law for energy conservation .

While, we can make another simple modification with the Demon. Choose to chain each $\lambda$ slice of Euclidean state with different energy which has achieve saturation of the entanglement entropic bound from unitarity (\ref{Unitary}), the evolution by the demon
\bea
\frac{dE}{d\lambda}=\braket{H_\partial}_\lambda-\braket{H_\partial}_{\lambda_0}
\eea 
labeled by $\lambda$, instead of the real physical time. (For example, we can let the Demon stops his work for some time, and continue latter. It doesn't change the whole story). The outgoing and incoming mode is at quasi-equilibrium, so we don't need to totally cancel ingoing part, so $T_{uu}\neq0$ and therefore the Gauss law is kept.

To see the entropic aspect, for each slice, the entanglement entropy achieve its maximal for the equilibrium. When we stop the demon at any time, the entanglement entropy will just stays its maximal bound, since it is enough real time evolution for the the system to scrambling. As the figure showing our proposer for the evolution in Figure \ref{EuclEvol}.

Let's get back to the massive graviton problem.  The crucial point to avoid massive graviton, is that we only get vanishing stress tensor, since we don't have real time $t$ rather a manipulated parameter $\lambda$. Our opinion is that such setup is actually working under quasi-static parameter rather than real time if we want the system evolving in equilibrium.

With the demon facility near the boundary controling the Dyson's sphere, the single boundary is still there, without actually ruining the gravitational Gauss law, so the massive graviton problem could be avoided.

\subsection{Reinterpret the Page Curve with Maxwell's Demon}

Though pure state in QG of closed universe can not be split like local QFT,  what is the meaning of Page curve? 

From the above Dyson's sphere, we use the demon, to reinterpret different entropy curves:
\begin{itemize}
   \item If we omit the hidden information of the demon, we find information loss with thermal entropy continue to increasing due to Hawking's calculation of radiation. This is a break of unitarity and leads to information loss.
   \item If we regard the demon and information as outside of QG Hilbert space, we get the original Page curve, which is pivoted at the Page time. This could be what the non-gravitating bath means, with gravitational subsystem not unitary but the whole system unitary.
   \item  When adding back the un-notice Demon to the system, the holography get recovered together with a trivial Page Curve in the closed Quantum Gravity system.

\end{itemize}

\begin{figure}[hbt]
		\centering
  \includegraphics[width=5in]{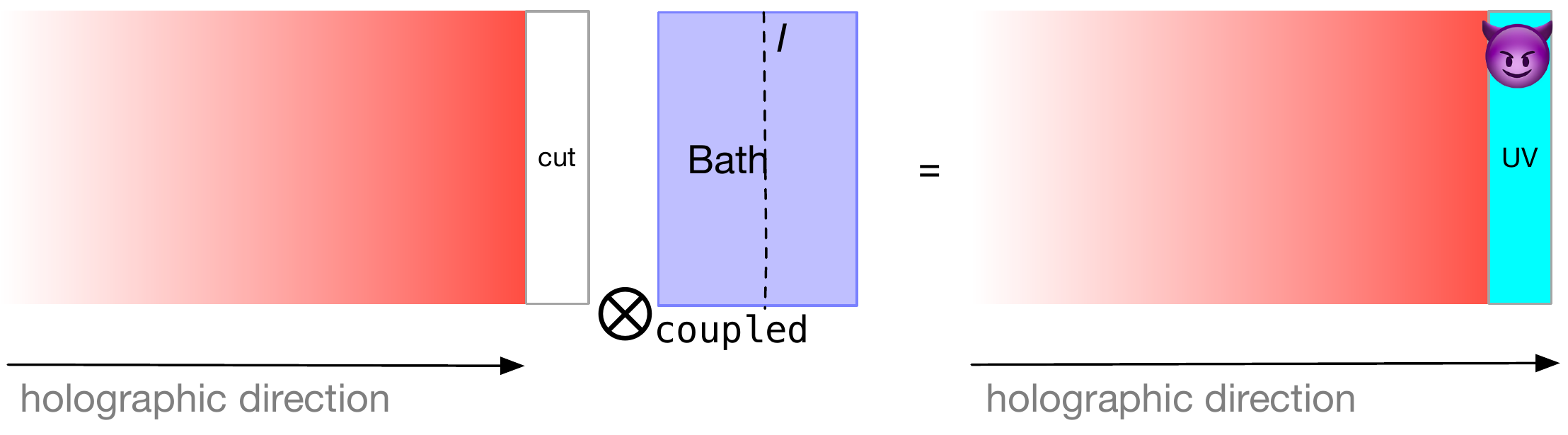}
  \caption{From an open system coupled to external bath, we propose it could be imitated by a close system with a felicity like Maxwell's Demon in the UV part. Thus, the process such as evaporation, low energy effective gravity coupled to external system should be able to be realized by the high energy facility which seems to violate the original evolving tendency.}
  \label{Demoneq}
\end{figure}

We know the UV/IR property of the renormalization of AdS/CFT \cite{Susskind:1998dq,deBoer:1999tgo}, that the near boundary region beyond the Dyson's sphere is also the UV part of quantum gravity.
When the inside observer deliberately ignore this part of UV gravitating region outside of his reach, the evaporating black hole state for closed QG system behaves as if coupled to non-gravitating bath, in semi-classical effective gravity theory.  We here suggest in Figure \ref{Demoneq} again that these two pictures are equivalent.


\subsection{How about Early time?}
How to extend the above analyze to the early time before the black hole state gets maximal-entangled? With the holographic entanglement entropy increase
\bea
\delta_\lambda S(R)>0
\eea
could the external force along radial direction is attractive and inward as
\bea
F_{ex}>0,\, F_{g}<0 ?
\eea

Although at  early time, we don't have the exact same condition to build the emergent gravity theory because
\begin{enumerate}
  \item the modular Hamiltonian is not $\partial_t$
  \item there is no area contribution to $S_{gen}$ for QES is trivial.
\end{enumerate}

Here the entanglement entropy is thermal entropy of the radiation $\H_R$ not of black hole $\H_{bh}$. Now, we suggest an interesting hypothesis:
\begin{quotation}
  \textit{Emergent gravity is actually responding to the variation of holographic entanglement entropy $\delta S_{vn}$, in Euclidean evolution picture with the demon/bath setup.}
\end{quotation}
Based on it, while at late time the radiation tends to get out, at early time, $\delta S(R)>0$ predicts the direction of tendency towards keeping the radiation not getting out: inward and attractive
\bea
F_g\propto -\delta_{\lambda}S_{vn}<0
\eea
 which is along the same direction of gravitational force.
Thus, comparing to the late time, the direction of evaporating out of the system is against the tendency: a draining out force should be executed by the Maxwell's demon to get the radiation out of system
\bea
F_{ex}>0\,.
\eea

\section{Summary and discussion}

 Radiating energy from gravitating system into external non-gravitating bath could mean extra external force.  
  We demonstrate the similarity between the heat flux $dQ$ into the bath and $F_{ex}$ that equilibrates the emergent gravitational force $F_g$.
  This effort is in a triumph to identify the fine-grained entropy variation in open systems, to an emergent effect in close systems equipped with some microscopic felicity in UV part.

To tell the gravitational tendency from thermodynamic tendency, the amount of entropy $S(M,m)$ is closely associated with the gravitational attraction, and the entropic gradient $\nabla S$ due to the locally isoenergic process shows the gravitational falling tendency. We have only observed the decreasing direction of the variation of the Casini-Bekenstein bound $\delta \Delta S$ is along the falling tendency during quasi-static. Besides, we also  update a better holographic meaning of this amount of entropic variation, from \cite{Chandra:2022fwi}, as the area variation of the apparent horizon, to the maximal of coarse-grained entropy.  Remind that here we are working on maximal entangled state of perturbing TFD state instead, we argue the extremal surface $\gamma$ is also varied in such amount. This approach may ultimately enable us to unify various natural tendency from the time arrow problem.  So far, we include the coarse-graining arrow and 2nd law tendency in the role of Maxwell's demon to make the processes in subsystem reversible.

Then, we interpret extracting energy and information out of the gravitating system as the evolution of Euclidean states in the bath setup, not only for the QES calculation of Page curve of Hawking evaporation, but also generally for the emergent gravitational force balanced with the external force.
Switching the situation from adiabatic process of close system to quasi-static process in  open system, we analogize perturbing the state in such processes to the evolution reversible heat engine, where one can use $F_{ex}$ and $\nabla S$ in equilibrating  to show the opposite of evolving direction of the adiabatic one,  and we suggest it also works in gravitational situation.
 Thus the falling tendency is simple  releasing heat into the bath by decreasing entropy bound in the gravitational subsystem, and $\bold F_{ex}$ completely transforms the gravitational potential to heat flux into the bath, where the direction is $dQ=d\Delta S>0$. 
When we withdraw such external force, the free-falling begins and the system evolves in coarse-graining.

Move up the exploration of tendency from gravitational attraction to the Hawking radiation, we discuss the possibility that bath setup is violating the original gravitational tendency in the black hole subsystem since $\delta S_{vn}\neq0$. The direction of entropy bound variation $\delta \Delta S$, could indicate the tendency in the late time, which is the island period.  Once the evaporation is equilibrated, one can form the second law view of the Page curve,
 to show the tendency of along the direction of radiating out the energy. We suggest it could be repulsive tendency at late time, so evolving direction is radiating out, such that $F_{ex}<0$ is inward and trying to go against the evaporation out. And we exhibit through the Dyson's sphere for the close system for the black hole to evaporating out but the whole system may still be kept pure.

By recalling the role of Maxwell's Demon, we reinterpret the relation among three curves: Hawking's calculation, the pivoted Page curve and the trivial Page curve. 
The pivoting of Page curve, could be also emergent, and co-exists with the trivial Page curve. We assign the new role to the historical Maxwell's Demon to help explore the mechanism of varying extremal surface and fine-grained entropy. To the black hole information paradox, the information may never get loss in the thermal Hawking radiation but hidden by canceling part of gravitating region, canceling the attraction between radiation.

The non-gravitational bath could have more deep reason for the Page curve to be emergent with the QES variation for the holographic entanglement entropy and entanglement wedge. In the close universe, our question is that could we build physics like the bath? Could it be some microscopic quantum Maxwell's Demon?
 A pity is that, we haven't shown the relevance to several interesting topics and deep question, such as emergence of radial direction, the island region, replica wormhole and other topics such as baby universe so far, but only use the conclusion of QES calculation for the $\delta S_{vn}$.

Besides, we suggest such demon could be the key to understand the Splitting Problem hanging behind the island progress. For Hilbert space irreducible as Type II or III von Neumann algebra \cite{Witten:2018lha}, the quantum gravity is the reason for meaningful $S_{gen}$ in (\ref{Sgen}) for finite dimensional Hilbert space.
The resemblance to the Maxwell's demon, suggests the clue to reveal the deep reason of the semi-classical split in the real word.

\noindent\textbf{Acknowledgments} \\
YA  would be grateful for useful discussion with Yu-Sen An, Peng Cheng, Shuwen Fu Shan-Ming Ruan, Jiarui Sun, Jia Tian, Zhenbin Yang, and support from the members including Anzhong Wang, Qiang Wu, Shaoyu Yin, Tao Zhu of the UCGWP group in Zhejiang University of Technology. Especially, we would thank Yang Zhou for questioning the 2nd Law direction, Edward Witten for a question if hidden degrees of freedom are involved.  This work is supported by the National Natural Science Foundation of China under Grant No. 12275238, No. 11975203, No. 11675143. YA is also seeking for visiting chance and potential cooperation.
\bibliographystyle{JHEP} 
\bibliography{EmergedGrav}

\end{document}